\documentclass[epj]{svjour}
\usepackage{amsmath}
\usepackage{amssymb}
\usepackage{epsfig}

\newlength{\feynwidth} 

\newcommand{\La}{{\Lambda}}
\newcommand{\Si}{{\Sigma}}

\newcommand{\be}{\begin{eqnarray}}
\newcommand{\ee}{\end{eqnarray}}

\begin{document} 
\title{In-medium properties of a {\boldmath$\Xi N$} interaction derived from chiral effective
field theory} 
\titlerunning{In-medium properties of a $\Xi N$ interaction}

\author{J. Haidenbauer$^{1}$ and U.-G. Mei\ss ner$^{2,1,3}$}
\authorrunning{J. Haidenbauer and U.-G. Mei\ss ner}

\institute{
$^1$Institute for Advanced Simulation, 
Institut f\"ur Kernphysik (Theorie) and J\"ulich Center for 
Hadron Physics, Forschungszentrum J\"ulich, D-52425 J\"ulich, Germany \\
$^2$Helmholtz Institut f\"ur Strahlen- und Kernphysik and Bethe Center
for Theoretical Physics, Universit\"at Bonn, D-53115 Bonn, Germany\\
$^3$Tbilisi State  University,  0186 Tbilisi, Georgia
}
\date{Received: date / Revised version: date}

\abstract{
The in-medium properties of a baryon-baryon interaction for the strangeness $S=-2$ 
sector derived within chiral effective field theory up to next-to-leading order
are investigated. 
The considered $\Xi N$ interaction is in line with available empirical constraints 
on the $\La\La$ $S$-wave scattering length and on $\Xi N$ elastic and inelastic 
cross sections. In particular, there are no near-threshold bound states in the 
$\Xi N$ system. 
A conventional $G$-matrix calculation for this interaction is performed and reveals 
that the single-particle potential of the $\Xi$-hyperon in nuclear matter 
is moderately attractive as suggested by recent experimental evidence for 
the existence of $\Xi$-hypernuclei. 
}
%

\maketitle

\section{Introduction}

A few years ago a hyperon-nucleon ($YN$) interaction has been derived up to next-to-leading 
order (NLO) in chiral effective field theory (EFT) by the J\"ulich-Bonn-Munich group \cite{Haidenbauer:2013}  
and soon afterwards an extension to baryon-baryon ($BB$) systems with strangeness $S=-2$ 
($\La\La$, $\Si\Si$, $\Xi N$) \cite{Haidenbauer:2015} was presented.
An excellent description of available $\Lambda N$ and $\Sigma N$ scattering data
could be achieved at NLO \cite{Haidenbauer:2013}. 
With regard to the $S=-2$ sector, the situation was much less conclusive simply
because of the scarcity of pertinent scattering data for the $\La\La$,
$\Si\Si$ and $\Xi N $ systems. Indeed, there are just a few data
points and upper bounds for the $\Xi N$ elastic and inelastic cross sections
\cite{Ahn:1997,Aoki:1998,Ahn:2006} that put constraints on the corresponding
interactions. Those constraints were met by the interaction proposed in
Ref.~\cite{Haidenbauer:2015}. However, it was also admitted that a
quantitative determination of the $S=-2$ $BB$ interaction and, thus, of key 
quantities like scattering lengths was not possible due to the lack of more
stringent experimental information. 

In the present paper we revisit the $\Xi N$ interaction within chiral EFT. 
Though, unfortunately, there is no new information on the elementary $\Xi N$ observables,
other developements over the last few years make this worthwhile. 
First, now there are stronger indications that bound $\Xi$-hypernuclei could indeed exist.
Specifically, evidence for deeply bound states in systems like $\Xi^-$-$^{\,14}$N \cite{Nakazawa:2015} 
(Kiso event) or $\Xi^-$-$^{\,11}$B \cite{Nagae:2017,Nagae:2018} have been
reported. 
%
Second, there is still an ongoing discussion about the role that hyperons play
for the size and stability of neutron stars. This concerns first of all 
the $\Lambda$ hyperon, but also the $\Xi$   
\cite{Weissenborn:2012,Weissenborn:2012a,Katayama:2015,Vidana:2016,Yamamoto:2016,Tolos:2017}. 
Third, the possibility to extract information on the $\Xi N$ interaction from
studying the pertinent correlations in heavy-ion collisions and/or high energetic 
$pp$ collisions might become feasible soon
\cite{Hatsuda:2017,Fabbietti:2018,Haidenbauer:2018}.
Finally, results from lattice QCD simulations for the $\Xi N$ interaction have become 
available for 
masses very close to the physical point \cite{Sasaki:2017,Sasaki:2018,Inoue:2018}.  

The results presented in Ref.~\cite{Haidenbauer:2015} suggest that the $\Xi N$ 
interaction has to be relatively weak in order to be in accordance with the available 
empirical information. In particular, the published values and upper bounds for the 
$\Xi^- p$ elastic and inelastic cross sections \cite{Aoki:1998,Ahn:2006} 
practically rule out a somewhat stronger attractive $\Xi N$ force and, specifically, 
disfavor any near-threshold deuteron-like bound states in that system that were
predicted by various phenomenological potentials in the past
\cite{Rijken:2010,Rijken:2015}. 
However, it remained open in how far such a very weakly attractive $\Xi N$ 
interaction like that in Ref.~\cite{Haidenbauer:2015} 
is compatible with in-medium properties deduced from studies of 
$(K^-,K^+)$ reactions on nuclei in the past \cite{Khaustov,Fukuda:1998} 
or with the aforementioned experimental evidence for possible deeply 
bound $\Xi$-hypernuclei \cite{Nakazawa:2015,Nagae:2017}.

We try to shed some light on this issue in the present work by investigating the 
properties of the $\Xi N$ interaction presented in Ref.~\cite{Haidenbauer:2015} 
in nuclear matter. Specifically, we report 
results for the single particle (s.p.) potential of the $\Xi$ hyperon in nuclear 
matter obtained in a conventional $G$-matrix calculation based on the standard 
(gap) choice for the intermediate spectrum. 
It turns out that the $\Xi N$ interaction proposed in Ref.~\cite{Haidenbauer:2015},
motivated primarily by the intention to show that a $BB$ interaction 
which fulfills all experimental constraints on $\La\La$ and $\Xi N$ scattering 
can be constructed within chiral EFT, is too repulsive in the medium. 
However, we will also demonstrate that an updated $\Xi N$ interaction can be 
established that fulfills likewise all experimental constraints on $\Xi N$ 
but yields a moderately attractive $\Xi$-nuclear potential.  

The paper is structured in the following way: 
In sect. 2 a summary of the employed formalism is provided. We briefly 
describe the ingredients and the construction of the $\Xi N$ interaction within 
chiral EFT and the equations used for calculating the nuclear matter properties. 
In sect. 3 we review the results for $\Xi N$ scattering of our initial 
$BB$ interaction \cite{Haidenbauer:2015} and we discuss the properties of
an updated version that is proposed in the present study. 
Subsequently we report results for the pertinent in-medium properties.
The paper ends with a brief summary. 

\section{Formalism}
\subsection{The {\boldmath$YN$} interaction in chiral EFT}

Details of the derivation of the baryon-baryon potentials for the 
strangeness sector within SU(3) chiral EFT can be be found in 
Refs.~\cite{Haidenbauer:2013,Haidenbauer:2015,Polinder:2006,Polinder:2007}, 
see also Ref.~\cite{Petschauer:2013}. Up to NLO the potentials consist of $BB$ contact terms 
without derivatives and with two derivatives, together with contributions from
one-pseudoscalar-meson exchanges and from (irreducible) two-pseudoscalar-meson exchanges.
On the one hand, the contributions from pseudoscalar-meson exchanges ($\pi$, $\eta$, $K$) are completely 
fixed by the assumed SU(3) flavor symmetry. On the other hand, the strength parameters associated with 
the contact terms, the so-called low-energy constants (LECs),
need to be determined in a fit to data. How this is done is described in detail in
Ref.~\cite{Haidenbauer:2013} for the $\La N$ and $\Si N$ channels and in
Ref.~\cite{Haidenbauer:2015} for the $BB$ interaction in the strangeness $S=-2$ sector. 
In practice, SU(3) flavor symmetry is also used as guideline for the contact terms because 
this allows one to reduce significantly the number of independent LECs that can contribute.
However, it turned out that a consistent description of, for example, $NN$ phase shifts 
and $\Si N$ scattering data is not possible under the assumption of strict SU(3)
symmetry for the contact terms. Because of that, we did not insist on rigorous SU(3) 
symmetry constraints for the employed LECs, when considering $NN$ and $YN$ or $YN$ and 
$YY$ \cite{HMP:2015,Haidenbauer:2015}. 
In any case this procedure is fully in line with SU(3) chiral EFT at NLO,
because at that order symmetry breaking terms in the leading $S$-wave contact 
interactions arise naturally in the perturbative expansion due to meson mass 
insertions \cite{Petschauer:2013}. 

The chiral $BB$ potential for $S=-2$ up to NLO is described in Ref.~\cite{Haidenbauer:2015}
and we refer the reader to this paper for details. Here we only want to 
recall the generic form of the contact interactions because some of the 
pertinent LECs will be readjusted in the course of the present study. 
The contact potentials in the various channels read \cite{Haidenbauer:2013}
\begin{eqnarray}
\nonumber
V&=&\tilde C+C\,(p^2+p'^2) \ ({\rm for } \ S \ {\rm waves}), \\ 
V&=&C\,pp' \ ({\rm for } \ P \ {\rm waves}) \ ,
\label{eq:LEC}
\end{eqnarray}
where $p$ and $p'$ are the center-of-mass (c.m.) momenta in the final and initial state, respectively,
and $\tilde C$ and $C$ denote the aforementioned LECs that specify the strength
of the contact terms. 
SU(3) symmetry implies that the contact potential in a specific $BB$ channel
is given by a particular combination of a basic set of $\tilde C$'s and $C$'s 
corresponding to the irreducible SU(3) representations
$8\otimes 8 = 27 \oplus 10^* \oplus 10 \oplus 8_s \oplus 8_a \oplus 1$,
relevant for scattering of two octet baryons \cite{Swa63,Dover1991}.
The generalized Pauli principle implies that the interaction in partial waves like
the $^3S_1$, which are symmetric with regard to their spin-space component, 
is given by linear combinations of LECs corresponding
to antisymmetric SU(3) representations ($10^*$, $10$, $8_a$),
whereas those with antisymmetric spin-space part ($^1S_0$) receive
only contributions from symmetric representations ($27$, $8_s$, $1$).
The concrete relations are summarized, e.g., in Table~1 of Ref.~\cite{Haidenbauer:2015}. 
 
The potentials in the various channels are partial-wave projected \cite{Polinder:2006} 
and then the reaction amplitudes are obtained from the solution of a coupled-channels 
Lippmann-Schwinger (LS) equation: 
\begin{eqnarray}
&&T^{\ell''\ell',J}_{\nu''\nu'}(p'',p';\sqrt{s})=V^{\ell''\ell',J}_{\nu''\nu'}(p'',p') 
\nonumber\\
&+&\sum_{\ell,\nu}\int_0^\infty \frac{dpp^2}{(2\pi)^3} \, V^{\ell''\ell,J}_{\nu''\nu}(p'',p)
\frac{2\mu_{\nu}}{k_{\nu}^2-p^2+i\eta}\nonumber\\
&\times&T^{\ell\ell',J}_{\nu\nu',J}(p,p';\sqrt{s})\ . \label{LS} 
\end{eqnarray}
The label $\nu$ indicates the particle channels and the label $\ell$ the partial wave. $\mu_\nu$ 
is the pertinent reduced mass. The on-shell momentum in the intermediate state, $k_{\nu}$, is 
defined by $\sqrt{s}=\sqrt{m^2_{B_{1,\nu}}+k_{\nu}^2}+\sqrt{m^2_{B_{2,\nu}}+k_{\nu}^2}$. 
Relativistic kinematics is used for relating the laboratory energy $T_{{\rm lab}}$ of the hyperons 
to the c.m. momentum.

In the calculation of phase shifts we solve the LS equation in isospin basis. In this case
up to three baryon-baryon channels can couple. For evaluating observables the LS equation is 
solved in the particle basis, in order to incorporate the correct physical thresholds. 
Then there are 6 coupled channels for the $\Xi^- p$ system. 
The potentials in the LS equation are cut off with a regulator function, 
$f_R(\Lambda) = \exp\left[-\left(p'^4+p^4\right)/\Lambda^4\right]$,
in order to suppress high-momentum components \cite{Epe05}. Cutoff values in the range 
$\Lambda=500$ -- $650$ MeV are considered, 
similar to what was used for chiral $NN$ potentials \cite{Epe05}. 
Following the tradition, we present our results as bands which reflect the variation
with the cutoff. These indicate primarily the uncertainty due to the employed regularization
scheme. Note that the LECs in the $S=-2$ sector cannot be uniquely determined due to 
the lack of near-threshold data and, therefore, a genuine uncertainty for the 
predictions cannot be given at present. This will certainly change in the future, when more
data will become available.

\subsection{Nuclear matter properties}
\label{SSA}

The nuclear matter properties of the $\Xi$ hyperon are evaluated within the conventional 
Brueckner theory. We summarize below only the essential elements. A detailed description 
of the formalism can be found in
Refs.\,\cite{Reuber:1994,Hai14}, see also Ref.\,\cite{Vid00}.
We consider the  $\Xi$ hyperon with momentum ${\vec p}_\Xi$ in nuclear matter at density $\rho$. 
In order to determine the in-medium properties of the hyperon we employ the Brueckner reaction-matrix
formalism and calculate the $\Xi N$ reaction matrix $G_{\Xi N}$, defined by the
Bethe-Goldstone equation
\begin{eqnarray}
\nonumber
&&\langle \Xi N | G(\zeta) | \Xi N \rangle = \langle \Xi N | V | \Xi N \rangle 
\\
&&+ \sum_{\cal X} \ \langle \Xi N | V | {\cal X} \rangle \,
\langle {\cal X} | \frac{Q}{\zeta - H_0}|{\cal X} \rangle \, \langle {\cal X} | G(\zeta) | \Xi N \rangle , 
\label{Eq:G1}
\end{eqnarray}
with ${\cal X}$ = $\Xi N$, $\La\La$, $\La\Si$, $\Si\Si$.
Here, $Q$ denotes the Pauli projection operator  which excludes intermediate
$\Xi N$-states with the nucleon inside the Fermi sea.
The starting energy $\zeta$ for an initial $\Xi N$-state with momenta ${\vec p}_\Xi $ and
${\vec p}_N$ is given by
\begin{equation}
\zeta = E_\Xi (p_\Xi) + E_N (p_N),
\end{equation}
where the single-particle energy $E_\alpha (p_\alpha)$ ($\alpha = \Xi, N$)
includes not only the (nonrelativistic) kinetic energy and the baryon mass but
in addition the single-particle potential $U_\alpha (p_\alpha, \rho)$:
\begin{equation}
E_\alpha (p_\alpha) = m_\alpha + \frac{\vec p^{\,2}_\alpha}{2m_\alpha} + U_\alpha (p_\alpha,
\rho)\, .
\label{Eq:G2}
\end{equation}
The conventional 'gap-choice' for the intermediate-state spectrum is adopted.
The $\Xi$ single-particle potential  \\ {$U_\Xi(p_\Xi,\rho)$
is given by the following integral and sum over diagonal $\Xi N$ $G$-matrix elements:
\begin{equation}
U_\Xi(p_\Xi,
\rho) = \int\limits_{|\vec p_N|< k_F}  {d^3p_N\over (2\pi)^3}\, \rm{Tr}
\langle {\vec p}_\Xi ,{\vec p}_N | G_{\Xi N}(\zeta) | {\vec p}_\Xi, {\vec p}_N \rangle 
\,,
\label{Eq:G3}
\end{equation}
where  $\rm{Tr}$ denotes the trace in spin- and isospin-space.
Note that $\rho = 2k^3_F/3\pi^2$ for symmetric nuclear matter considered in the present
work, where $k_F$ denotes the Fermi momentum. 
Eqs.\,(\ref{Eq:G1}) and (\ref{Eq:G3}) are solved self-consistently in the standard way,
with  $U_\Xi(p_\Xi, \rho)$ appearing also in the starting energy $\zeta$. Like in 
Ref.~\cite{Haidenbauer:2017} the nucleon single-particle potential
$U_N(p_N,\rho)$ is taken from a calculation of nuclear matter employing a
phenomenological $NN$ potential. Specifically, we resort to results for the Argonne $v_{18}$
potential published in Ref.\,\cite{Isaule:2016}. 
As pointed out in Ref.~\cite{Reuber:1994}, calculations of hyperon potentials in nuclear 
matter using the gap-choice are not 
too sensitive to the details of $U_N(p_N,\rho)$. Indeed, the difference for $U_\Xi(0,\rho)$ 
using $U_N(p_N,\rho)$ from Ref.~\cite{Isaule:2016} or the parameterization utilized in
Ref.~\cite{Hai14} amounts to less than $1$~MeV at nuclear matter saturation
density $\rho= 0.17$~fm$^{-3}$ ($k_F=1.35$~fm$^{-1}$).

Since at this stage we are only interested in getting a qualitative insight, we refrain 
from a much more time-consuming calculation necessitated by the  so-called `continuous choice'  
\cite{Petschauer:2015}. 
A comparison of our leading-order (LO) results with the ones of Kohno~\cite{Kohno} 
suggests that the continuous choice lowers the $\Xi$-nuclear potential by roughly $3$~MeV at 
$k_F=1.35$~fm$^{-1}$. According to Yamamoto et al.~\cite{Yamamoto:2010} 
the difference (increase in attraction) could be, however, as large as $10$~MeV. 

\section{Results}
\subsection{{\boldmath$\Xi N$} scattering}
\label{SSR}

\begin{figure}[t]
\includegraphics*[width=6cm,angle=-90]{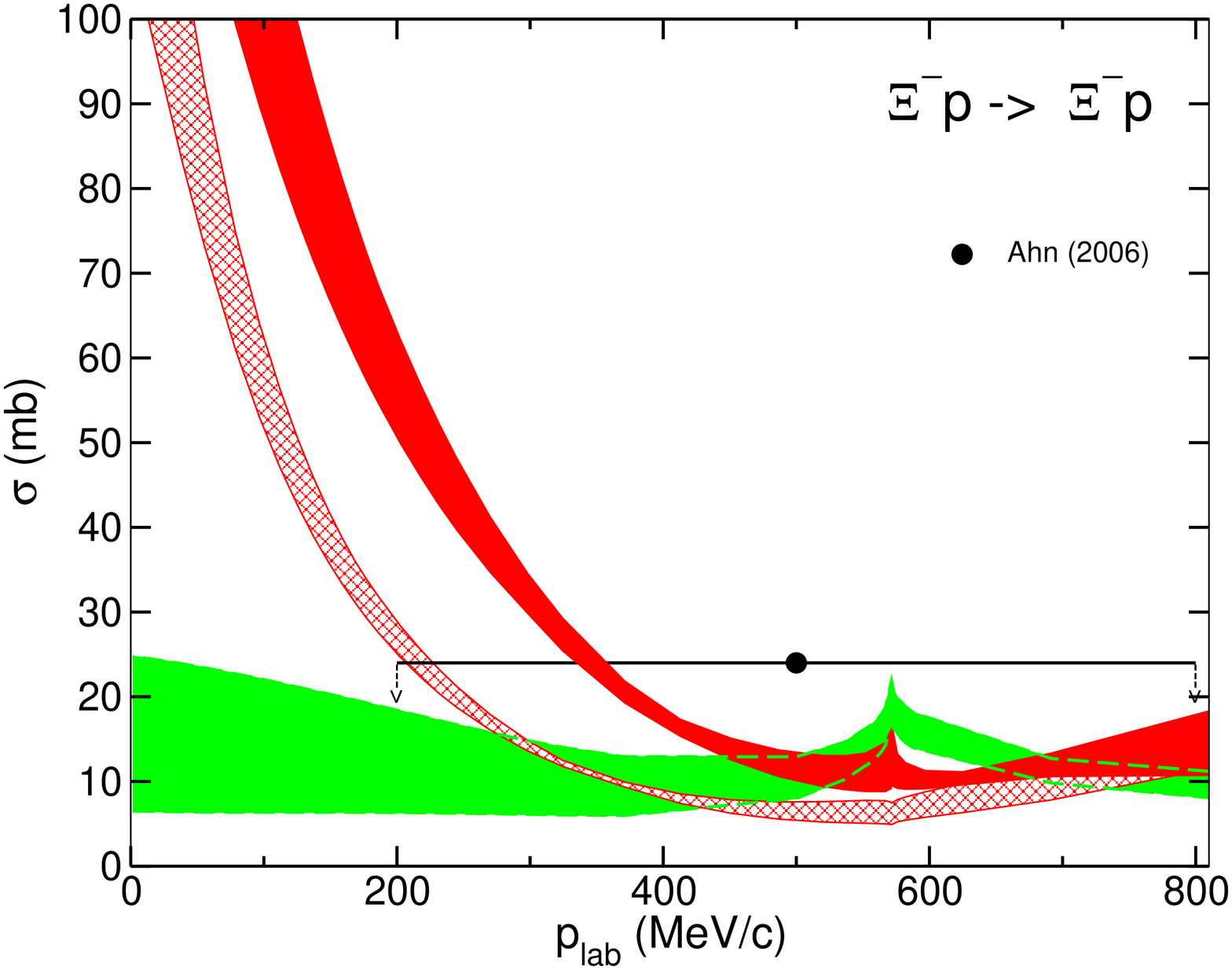}
\includegraphics*[width=6cm,angle=-90]{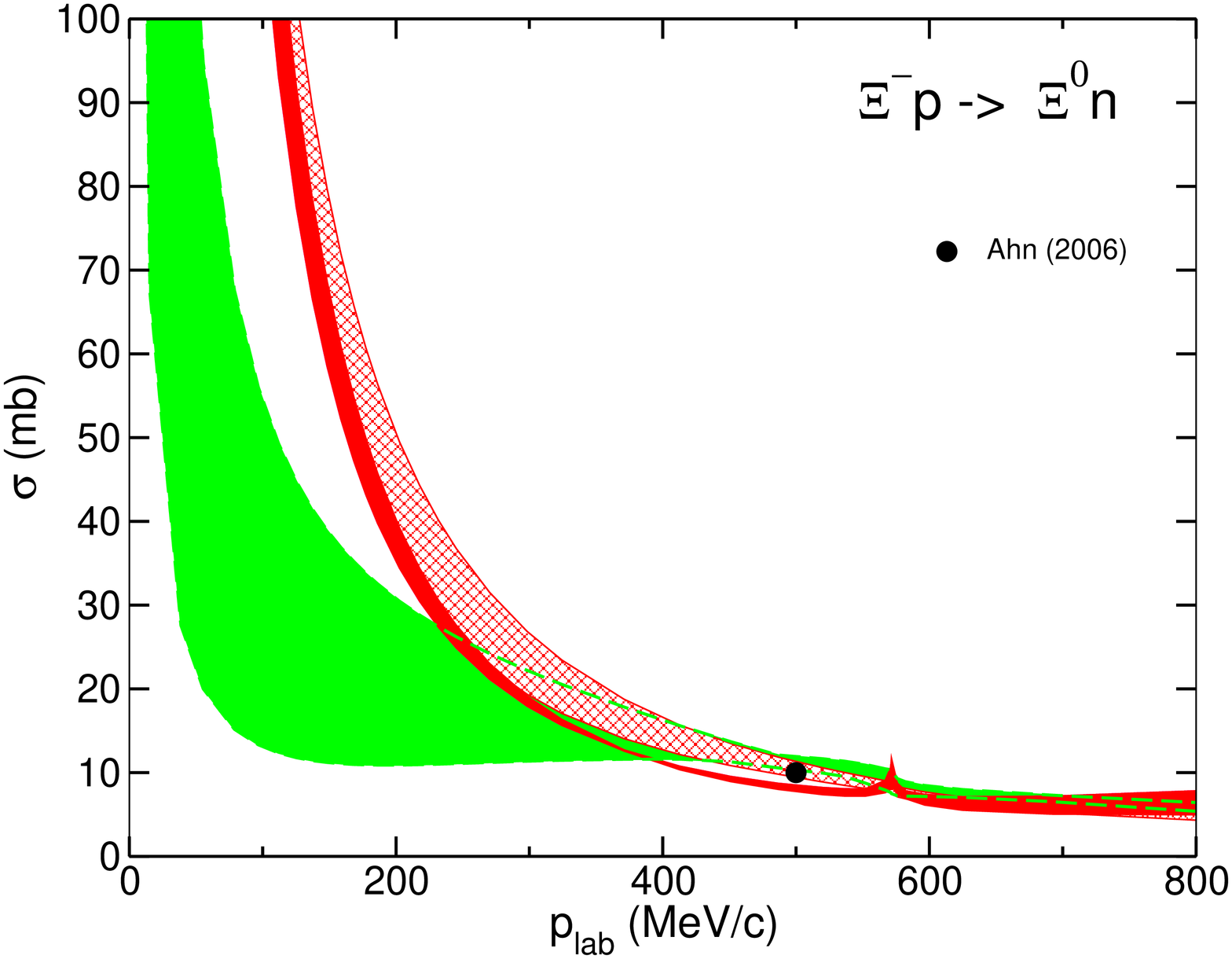}
\includegraphics*[width=6cm,angle=-90]{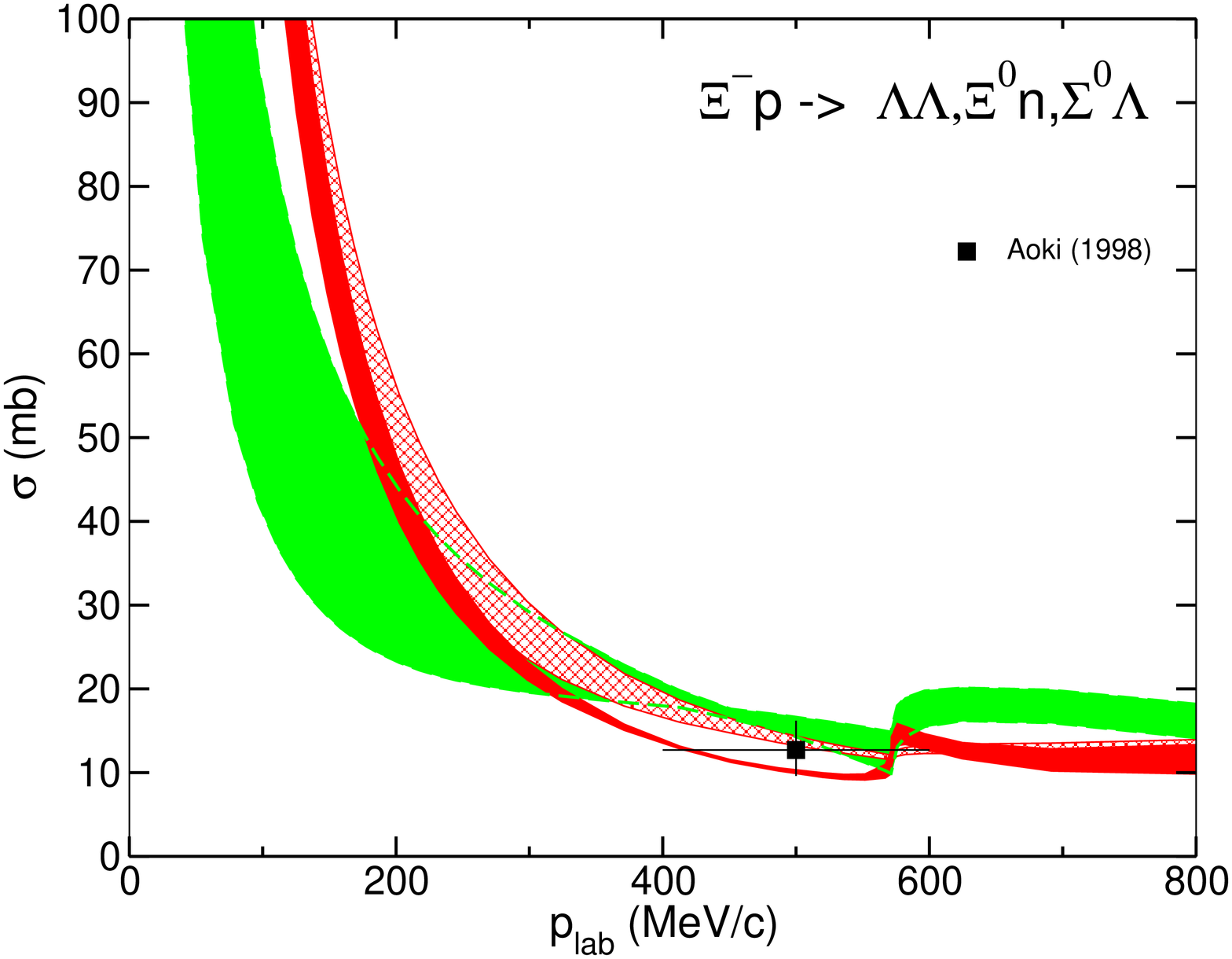}
\caption{$\Xi^- p$ induced cross sections. The black/red bands represent results 
at NLO, based on the new fit with readjusted LECs $\tilde C^{10^*}_{^3S_1}$ 
and $\tilde C^{10}_{^3S_1}$, see text. 
The hatched bands are results for the NLO interaction from 
Ref.~\cite{Haidenbauer:2015} while the grey/green bands are those from a LO 
calculation \cite{Polinder:2007}.  
Experiments are from Ahn et al.~\cite{Ahn:2006} and Aoki et al.~\cite{Aoki:1998}.
}
\vspace{-5mm}
\label{XS2}
\end{figure}

Our initial study of the baryon-baryon interaction in the strangeness $S=-2$ sector
was guided by the aim to meet all the (limited and mostly qualitative) experimental 
constraints on the $\La \La$ \cite{Gasparyan:2012,Haidenbauer:2015,Ohnishi:2016} 
and $\Xi N$ systems \cite{Ahn:1997,Aoki:1998,Ahn:2006}.
In practice this meant that we did not take over the employed LECs from our 
earlier $YN$ study \cite{Haidenbauer:2013} in all cases as it would be compulsory 
for strict SU(3) symmetry. Rather we exploited the freedem that within SU(3) 
chiral EFT at NLO symmetry breaking terms in the leading $S$-wave contact 
interactions already arise~\cite{Petschauer:2013}. Specifically, 
for the $^1S_0$ partial wave a study of the $pp$ and $\Sigma^+ p$ systems 
allowed us to quantify the amount of SU(3) symmetry breaking and then infer,
in line with SU(3) chiral EFT, the corresponding LEC ($\tilde C^{27}_{^1S_0}$) 
to be used for the $S=-2$ sector \cite{HMP:2015}. 
For the LECs in the $^3S_1$-$^3D_1$ partial wave such a procedure was not viable.
Therefore, and somewhat arbitrarily the value of $\tilde C^{8_a}_{^3S_1}$ was 
readjusted in the $S=-2$ sector, whereas the other ones 
($\tilde C^{10^*}_{^3S_1}$ and $\tilde C^{10}_{^3S_1}$) were taken over from
the $YN$ study \cite{Haidenbauer:2013}. Indeed readjusting that LEC allowed
to avoid a near-threshold bound state in the $I=0$ channel, which arises for the 
value of $\tilde C^{8_a}_{^3S_1}$ fixed in the fit to $YN$ data and which is the 
main reason why the $\Xi^- p$ cross section based on the LECs from our $YN$ 
study exceeds the upper limits set by experiments \cite{Ahn:2006} dramatically,  
see the extensive discussion in Ref.~\cite{Haidenbauer:2015}. 

As we will show below, the $\Xi N$ interaction fixed in that way leads to 
unrealistic in-medium properties. Specifically, the s.p. potential for the
$\Xi$ turns out to be much too repulsive. 
Therefore, in the present paper we seek for an alternative solution, where 
now we allow variations of the LECs $\tilde C^{10^*}_{^3S_1}$ 
and $\tilde C^{10}_{^3S_1}$ too. 
Our aim is to explore the possibility to establish a $\Xi N$ interaction
that still meets all the experimental constraints for $\La \La$ and $\Xi N$
scattering, but at the same time implies an attractive $\Xi$ s.p. potential. 
Thereby, the interaction in the $^1S_0$ partial wave is taken over from 
our previous work \cite{Haidenbauer:2015}. After all, our predictions for
$\La\La$ and $\Xi N \to \La\La$ are well in line with experiments and, 
moreover, they are supported by the latest (though still preliminary) results 
from lattice QCD simulations \cite{Sasaki:2017}. The latter is also
the case for the $\Xi N$ $^3S_1$-$^3D_1$  interaction in the isospin $I=0$ 
channel so that we simply take over the values for $\tilde C^{8_a}_{^3S_1}$ 
fixed in Ref.~\cite{Haidenbauer:2015}. 

Results for those $\Xi N$ channels where data are available are presented
in Fig.~\ref{XS2}. Here the hatched bands are the NLO results from
Ref.~\cite{Haidenbauer:2015} and the black/red bands those of the
new alternative solution. For illustration we include also predictions obtained
at leading order (LO) \cite{Polinder:2007}, cf. the light (green)
bands. As already said above, the results for the 
$\La\La \to \La\La$ and $\Xi^-p \to \La\La$ cross sections remain unchanged
and, therefore, are not reproduced here. 

\begin{figure}
\includegraphics*[width=6.2cm,angle=-90]{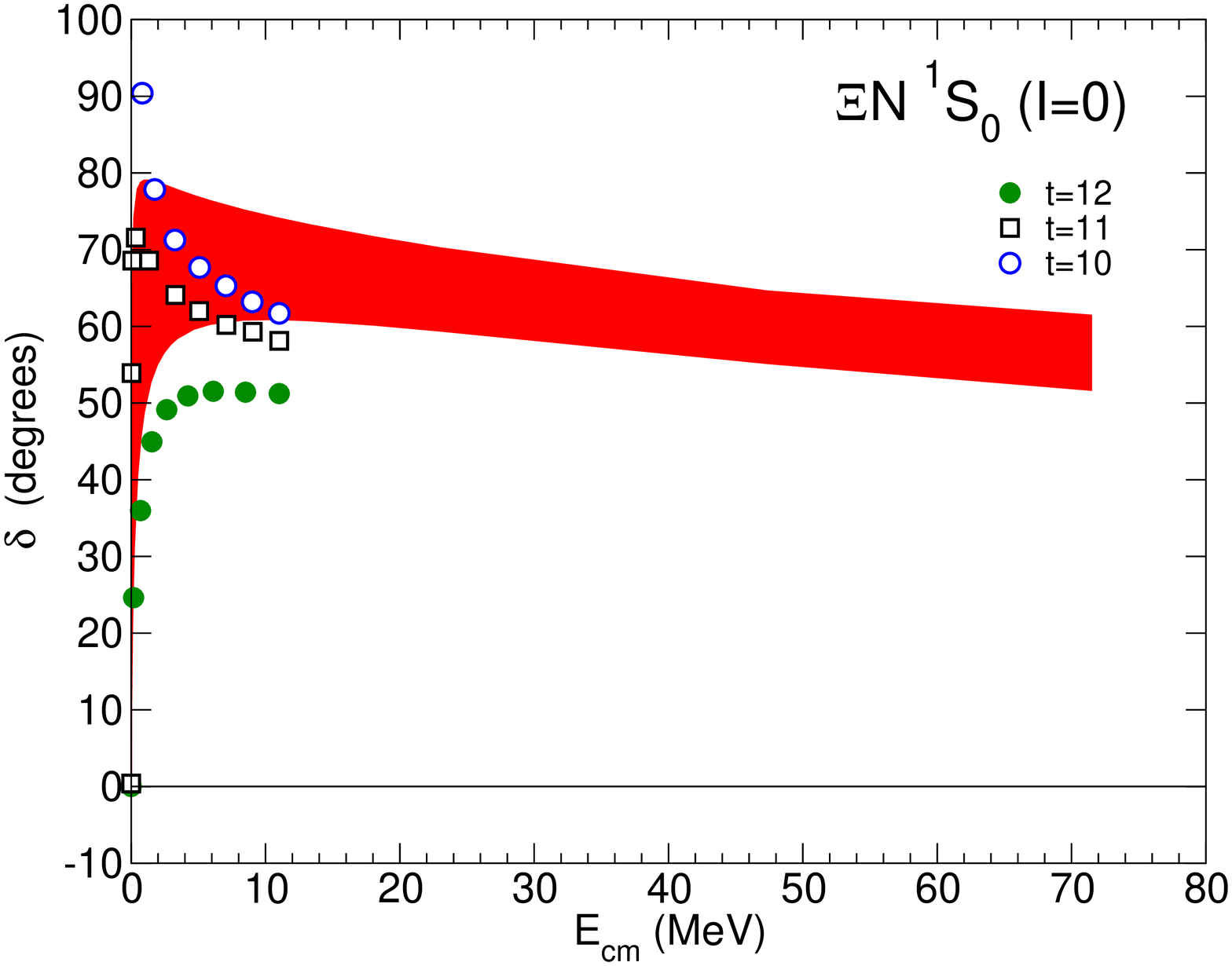}
\includegraphics*[width=6.2cm,angle=-90]{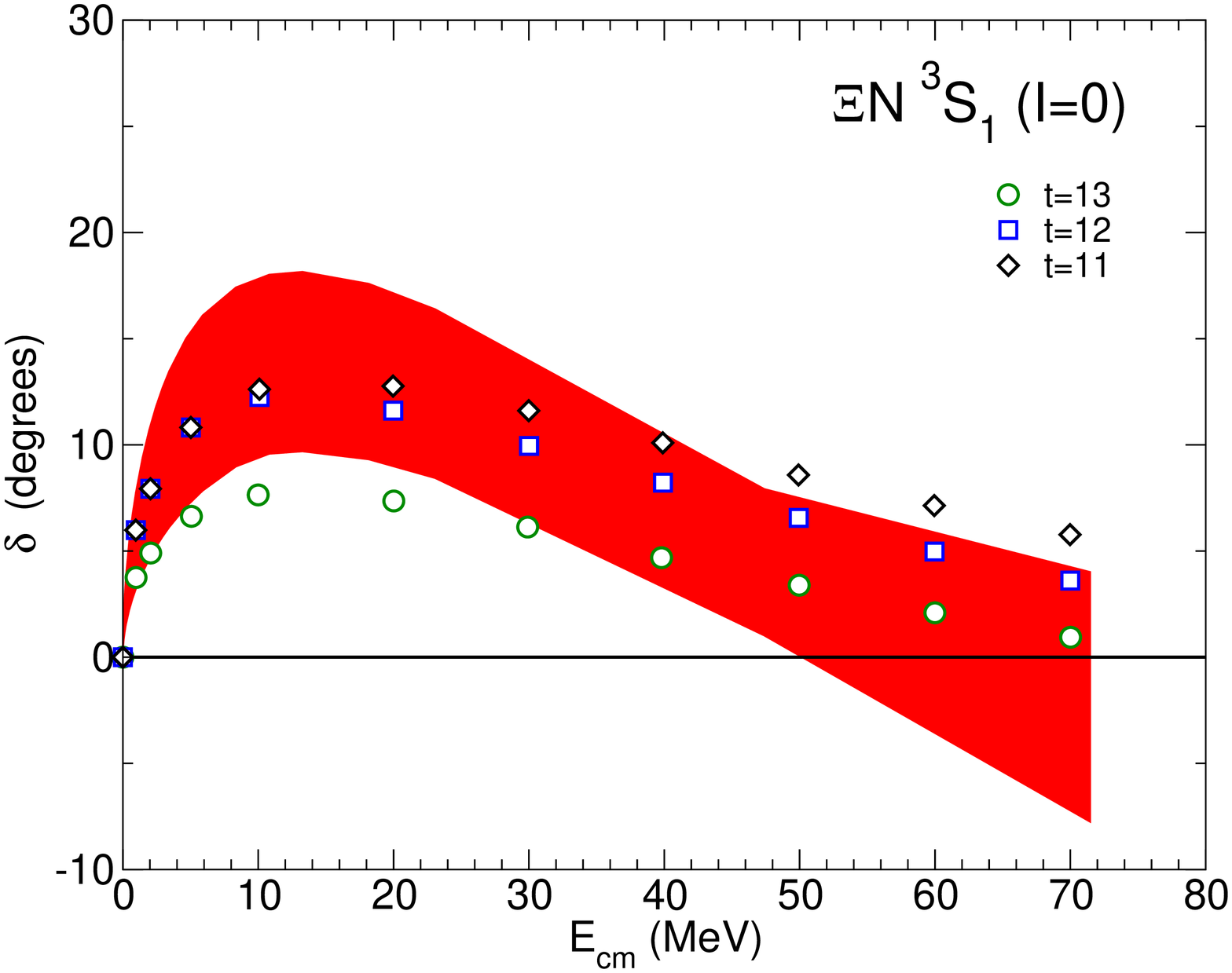}
\caption{$\Xi N$ isospin $I=0$ phase shifts from Ref.~\cite{Haidenbauer:2015}. 
The symbols indicate preliminary results from lattice QCD calculations by the 
HAL QCD collaboration for different sink-source time-separations $t$ \cite{Sasaki:2017}. 
}
\label{Ph2}
\end{figure}
\begin{figure}
\includegraphics*[width=6.2cm,angle=-90]{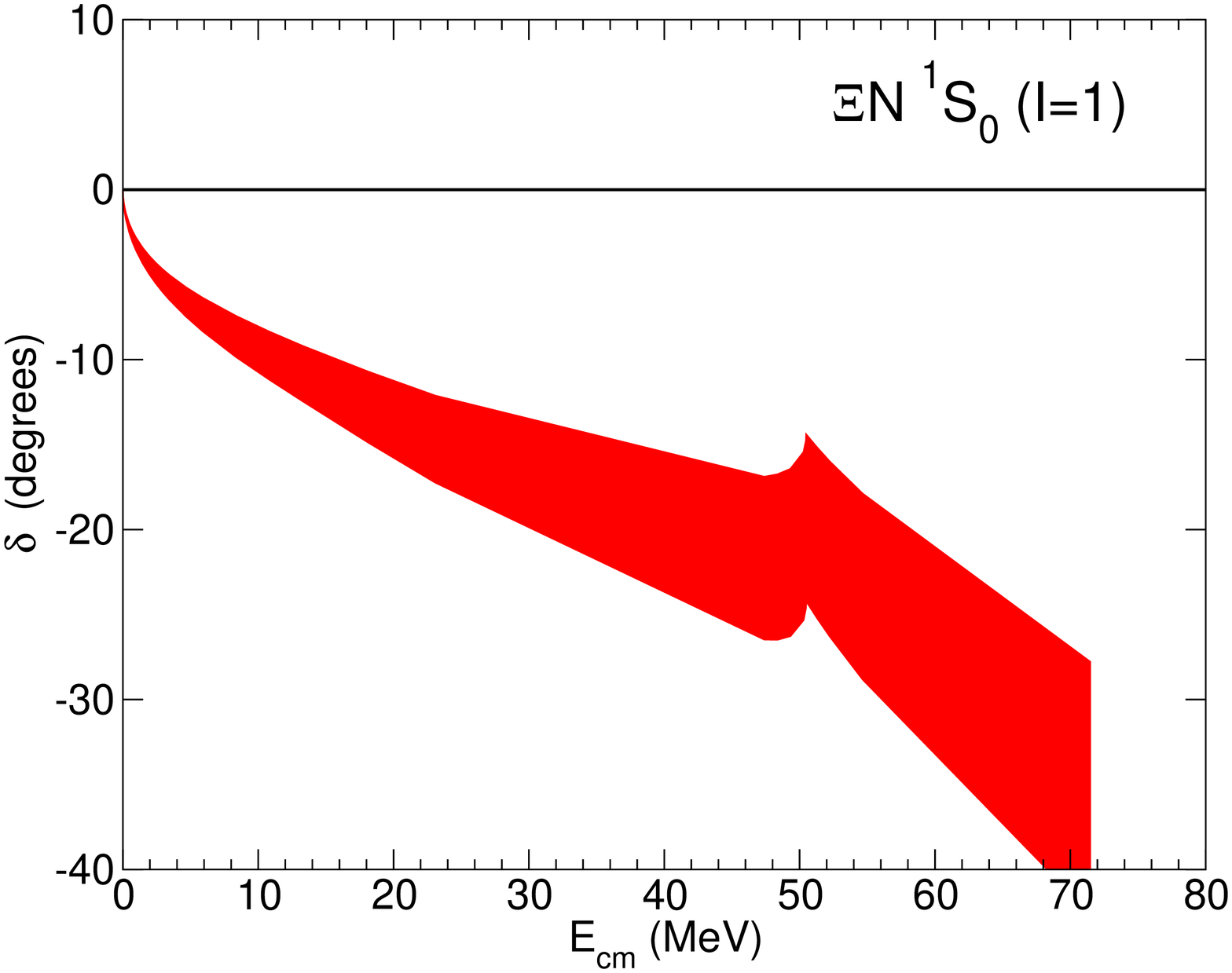}
\includegraphics*[width=6.2cm,angle=-90]{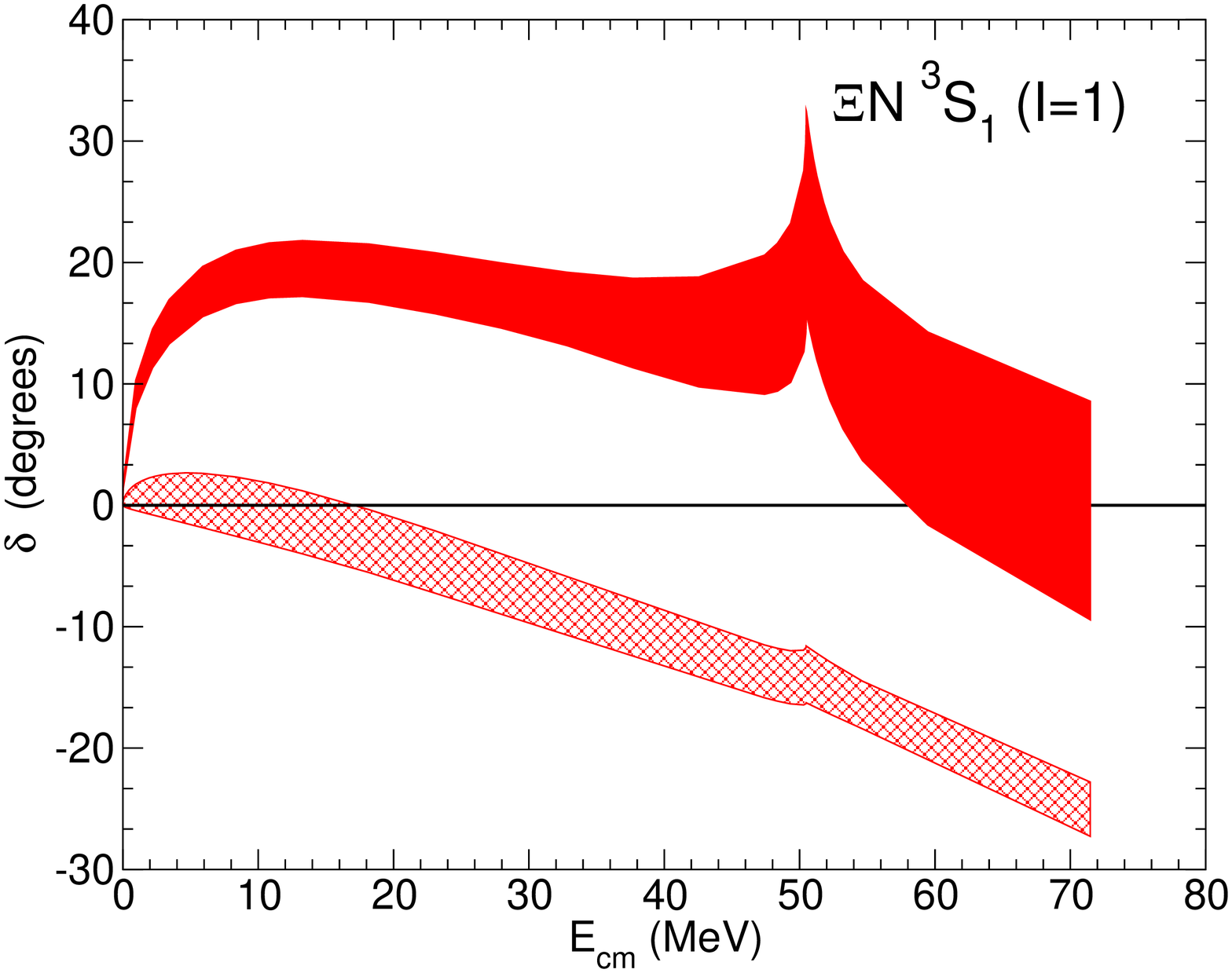}
\caption{$\Xi N$ isospin $I=1$ phase shifts. The hatched band are the 
$^3S_1$ phase shifts for the interaction presented in Ref.~\cite{Haidenbauer:2015}. 
The full band represents the one for the updated $\Xi N$ 
interaction considered in the present work. 
}
\label{Ph3}
\end{figure}

As visible in Fig.~\ref{XS2}, 
the main difference between the $\Xi N$ interaction from Ref.~\cite{Haidenbauer:2015}
and the new fit is that in the former the $\Xi^- p$ elastic cross section remains
strictly below the upper bound while now the limit provided by the experiment is 
fulfilled only in average over the given momentum range of $200 < p_{\rm lab} < 800$~MeV/c.
Both scenarios are, of course, consistent with the empirical findings 
\cite{Ahn:2006}. 
The phase shifts in the $\Xi N$ $S$-waves are summarized in Figs.~\ref{Ph2}
and \ref{Ph3}. 
For completeness we show here all $S$-waves though the alternative solution concerns 
only the $^3S_1$-$^3D_1$ partial wave with $I=1$. 
One can see that now the interaction in the latter partial wave is moderately attractive  
while it was basically repulsive in our previous work \cite{Haidenbauer:2015}. 
Interestingly, this attraction leads to a much more pronounced cusp effect at the opening of the 
$\La\Si$ channel, comparable to what happens in the $\La N$ case at the opening of 
the $\Si N$ channel \cite{Haidenbauer:2013,Machner:2013}.  

Table~\ref{ERE} provides a summary of the pertinent $S$-wave effective 
range parameters. Besides the ones of our chiral EFT interactions 
we included values for two phenomenological potential models from the 
literature, whose $G$-matrix results will serve us as benchmark in the 
discussion of in-medium properties below. 
The models in question are the Nijmegen ESC08c meson-exchange potential 
\cite{Rijken:2015} and the quark-model potential fss2 \cite{Fujiwara07}. 
Note that the large and positive value of $a_{^3S_1}$ for $I=1$ in case of 
the Nijmegen ESC08c potential indicates the presence of a bound state in 
this partial waves.
A more comprehensive overview of $\Xi N$ effective range parameters predicted 
over the years can be found in Ref.~\cite{Gasparyan:2012}. 

As already mentioned, recently lattice QCD results close to the physical point 
($M_\pi = 146$~MeV) have become available for the $S=-2$ sector from the HAL QCD 
Collaboration \cite{Sasaki:2017,Sasaki:2018}. 
The reported phase shifts for the $\Xi N$ $^1S_0$ and $^3S_1$ partial waves
with isospin $I=0$ are very similar to the ones predicted by our EFT
interaction earlier~\cite{Haidenbauer:2015}.  
This can be seen in Fig.~\ref{Ph2}, where the phases from the lattice are included for 
illustration. 
We want to emphasize, however, that the HAL QCD results are still preliminary. Also, 
we show here only the central values for different sink-source time-separations 
$t$ \cite{HALQCD:2012}. For the pertinent statistical errors see Ref.~\cite{Sasaki:2017}. 
In any case, one interesting aspect is that the present lattice results support the 
possible existence of a virtual state in the $I=0$ $^1S_0$ partial wave very 
close to the $\Xi N$ threshold, predicted in \cite{Haidenbauer:2015}, 
which can be considered as a remnant of the $H$-dibaryon 
\cite{Haidenbauer:2011a,Haidenbauer:2011b}. It is reflected in large values
of the corresponding $\Xi N$ phase shifts, see upper panel of Fig.~\ref{Ph2},
and large negative values of the scattering length, cf.~Table~\ref{ERE}. 
It is also worthwhile mentioning that the HAL QCD lattice simulations suggest 
values around $-0.65$~fm for $a^{I=0}_{^3S_1}$ \cite{Sasaki:2017}, which is
well within the range predicted by our NLO potential \cite{Haidenbauer:2015}. 

\begin{table*}[t]
\caption{$\Xi N$ scattering lengths and effective ranges (in fm) 
for the NLO potential for cutoffs $\Lambda=500-650$~MeV. Results for
the interaction considered in Ref.~\cite{Haidenbauer:2015} are
shown in brackets when different. Values for the Nijmegen ESC08c 
potential \cite{Rijken:2015} and
from the quark-model potential fss2 \cite{Fujiwara07} are included too. 
}
\label{ERE}
\vspace{0.2cm}
\centering
\renewcommand{\arraystretch}{1.5}
\begin{tabular}{|l||cc|cc||cc|cc|}
\hline
         &\multicolumn{4}{|c||}{$^1S_0$} & \multicolumn{4}{|c|}{$^3S_1$} \\
\hline
         &\multicolumn{2}{|c|}{$I=0$} & \multicolumn{2}{|c||}{$I=1$} 
         &\multicolumn{2}{|c|}{$I=0$} & \multicolumn{2}{|c|}{$I=1$} \\
\hline
{potential} & \multicolumn{2}{|c|}{$a$}  & $a$ & $r$ & $a$ & $r$ & $a$ & $r$ \\
\hline
NLO (500) &  $-7.71-\,i\,2.03$  &  & 0.37 & $-$4.80 & $-$0.33 & $-$6.86 & $-$1.17 & 3.44 \\
          &                   &  &      &       &       &      &(-0.20  &  35.6) \\
NLO (550) &  $-7.24-\,i\,20.79$ &  & 0.39 & $-$4.95 & $-$0.39 & $-$1.77 & $-$1.15 & 3.80 \\
          &                   &  &      &       &       &      &(-0.04  &  575) \\
NLO (600) & $-10.89-\,i\,14.91$ &  & 0.34 & $-$7.20 & $-$0.62 & 1.00 & $-$1.13 & 3.95 \\
          &                   &  &      &       &       &      &(0.02  &  1797) \\
NLO (650) &  $-8.14-\,i\,2.43$  &  & 0.31 & $-$9.16 & $-$0.85 & 1.42 & $-$0.90 & 4.27 \\
          &                   &  &      &       &       &      &(0.04  &  450) \\
\hline
\hline
ESC08c     &  &  & 0.58 & $-$2.52   & $-$5.36 & 1.43 & 4.91 & 0.53 \\
\hline
fss2       &  &  &      &         &  0.32 & $-$8.93 & $-$0.21 & 26.2 \\
\hline
\end{tabular}
\renewcommand{\arraystretch}{1.0}
\end{table*}

\begin{table}
\caption{New values of the LECs in the $^3S_1$ partial wave for the considered cutoffs $\Lambda$,
in the notation of Ref.~\cite{Haidenbauer:2015}. In addition $C^{8_s}_{^3P_0} = -0.5$ and 
$C^{8_s}_{^3P_2} = -0.15$ are used in the alternative NLO fit. 
The values for $\tilde C$ are in 10$^4$~GeV$^{-2}$, those for the $C$'s in 10$^4$~GeV$^{-4}$.
}
\label{LEC}
\vspace{0.2cm}
\centering
\renewcommand{\arraystretch}{1.5}
\begin{tabular}{|l|cccc|}
\hline
$\Lambda$  [MeV] & 500 & 550 & 600 & 650 \\
\hline
\ $\tilde C^{10^*}_{^3S_1}$ \ & \ 0.541 \ & \ 1.49 \ & \ 1.64 \ & \ 2.40 \ \\ 
$\tilde C^{10}_{^3S_1}$   & 0.011 & 0.05 & 0.62& 1.20 \\ 
\hline
\end{tabular}
\renewcommand{\arraystretch}{1.0}
\end{table}

\subsection{{\boldmath$\Xi$} in nuclear matter}

The properties of our $\Xi N$ interaction in nuclear matter are
documented in Table~\ref{UXI} and Fig.~\ref{kdep}. The table
summarizes the results for the $\Xi$ potential depth, 
$U_\Xi (p_\Xi = 0)$, evaluated at the saturation point of
nuclear matter, i.e. for $k_F = 1.35$~fm$^{-1}$, for the NLO interaction 
of Ref.~\cite{Haidenbauer:2015} and the updated interaction considered in 
the present study. For completeness we include 
also an exemplary result for our LO potential \cite{Polinder:2007}. 
In addition, results for two potentials from the literature are 
listed, namely for the Nijmegen ESC08c potential \cite{Rijken:2015}
and the quark-model based potential fss2 \cite{Fujiwara07}. 

There are indications for a moderately attractive in-medium $\Xi N$ interaction, 
as already pointed out in the Introduction. In particular, a strength
of $U_\Xi \approx -14$~MeV for the $\Xi$-nucleus potential is considered as 
the benchmark now \cite{Gal:2016}. This value was deduced from the initial 
analysis of the BNL-E885 measurement of the spectrum of the $(K^-,K^+)$ 
reaction on a $^{12}$C target based on a Woods-Saxon potential \cite{Khaustov}.
Since we calculate the $\Xi$-nuclear potential in infinite nuclear matter 
we should, however, not really compare our result directly with that 
figure. At least this is suggested by investigations in the literature
where $G$-matrix calculations similar to ours were presented and where, 
in addition, analyses of finite $\Xi$ systems were performed utilizing 
$\Xi$-nucleus potentials derived from those $G$-matrices.
For instance, in studies based on the Nijmegen ESC08 potentials \cite{Rijken:2015,Yamamoto:2016} 
predictions for $\Xi$-hypernuclei were reported which are apparently consistent with the 
indications of the BNL experiment \cite{Khaustov} and with the Kiso event \cite{Nakazawa:2015}. 
However, if we take exemplary the ESC08c model, the corresponding potential in nuclear matter, 
evaluated at saturation density, amounts to $-7$~MeV only \cite{Rijken:2015}.  
In the analysis by Kohno and Hashimoto \cite{Kohno:2010} (see also Ref.~\cite{Kohno:2009})
that is partly based on the quark-model potential fss2 \cite{Fujiwara07} the authors came 
likewise to the conclusion that the BNL data do not necessarily imply an attractive $\Xi$-nucleus 
potential in the order of $14$~MeV. Indeed, in this work it is concluded that an almost zero 
potential is preferable.

\begin{table*}[t]
\caption{Results for $\Xi$ in nuclear matter. Partial wave contributions to  
$U_\Xi(p_\Xi=0)$ (in MeV) at the Fermi momentum $k_F = 1.35$ fm$^{-1}$, 
for the LO \cite{Polinder:2007} and NLO interactions for various cutoffs. 
NLO$^*$ indicates the NLO interaction from Ref.~\cite{Haidenbauer:2015}. 
Results for the Nijmegen ESC08c potential are taken from Ref.~\cite{Rijken:2015}, 
those for the quark-model potential fss2 from Ref.~\cite{Kohno:2010}.
}
\label{UXI}
\vspace{0.2cm}
\centering
\renewcommand{\arraystretch}{1.5}
\begin{tabular}{|l|c|rrrrrr|c|c|}
\hline
\hline
potential & \ $I$ \  & \ $^1S_0$ \ & \ $^3S_1$ \ & \ $^3P_0$ \ & \ $^1P_1$ \ & \ $^3P_1$ \ & \ $^3P_2$ \ & \ $S$-waves \ & \ total \ \\
\hline
NLO (500) \ & 0 & $-$2.6 & $-$3.3  & $-$0.7 & $-$0.9 &  1.7 & $-$0.6 &&            \\
           & 1 & 12.7 & $-$11.8 &  1.4 &  1.5 & $-$0.3 & $-$2.5 &$-$5.0& $-$5.5       \\
NLO (550) & 0 & $-$2.9 & $-$3.1  & $-$0.7 & $-$0.9 &  1.7 & $-$0.6 &&            \\
          & 1 & 12.4 &  $-$9.5 &  1.3 &  1.4 & $-$0.4 & $-$2.5 &$-$3.1& $-$3.8       \\
NLO (600) & 0 & $-$2.9 & $-$3.8  & $-$0.8 & $-$0.9 &  1.7 & $-$0.6 &&            \\
           & 1 & 10.4 & $-$7.0  &  1.3 &  1.4 & $-$0.5 & $-$2.6 &$-$3.3& $-$4.3       \\
NLO (650) & 0 & $-$2.7 & $-$4.8  & $-$0.8 & $-$1.0 &  1.6 & $-$0.6 &&            \\
           & 1 &  9.1 & $-$4.1  &  1.2 &  1.3 & $-$0.6 & $-$2.7 &$-$2.5& $-$4.1       \\
\hline
NLO$^*$(500) & 0 & $-$3.1 & $-$3.3   & $-$0.5& $-$0.9 &   1.7 &  0.1 &&            \\
          & 1 & 11.4 &   9.1  &  3.1&  1.5 &  $-$0.4 &  4.4 &14.1& 23.1       \\
NLO$^*$(550) & 0 & $-$3.6 & $-$3.2   & $-$0.5& $-$0.9 &   1.7 &  0.1 &&            \\
          & 1 & 10.9 &  16.0  &  2.7&  1.4 &  $-$0.4 &  3.8 &20.1& 27.7       \\
NLO$^*$(600) & 0 & $-$3.5 & $-$4.1   & $-$0.6& $-$1.0 &   1.7 &  0.0 &&            \\
          & 1 &  9.0 &  18.0  &  2.4&  1.4 &  $-$0.5 &  3.2 &19.4& 26.0       \\
NLO$^*$(650) & 0 & $-$3.2 & $-$5.2   & $-$0.6& $-$1.0 &   1.6 & $-$0.1 &&            \\
          & 1 &  8.0 &  17.2  &  2.2&  1.3 &  $-$0.7 &  2.8 &17.0& 22.4       \\
\hline
 LO (600 ) & 0 & $-$1.2 & $-$4.9  & $-$1.1& $-$0.5 &   1.7 & $-$0.3 &&            \\
           & 1 &  6.8 &  9.8  &  1.1&  0.9 &  $-$3.1 &  0.2 &10.5&  9.3       \\
\hline
\hline
ESC08c & 0 & 1.4  & $-$8.0      & 1.8 & $-$0.3 &  1.4  & $-$2.1 && \\
       & 1 & 10.7 & $-$11.1     & 0.7 & 1.1 &  $-$2.6 & $-$0.0 &$-$7.0& $-$7.0  \\
\hline
fss2   &   &      &      &      &     &      &      &&  $-$1.5  \\
\hline
\hline
\end{tabular}
\renewcommand{\arraystretch}{1.0}
\end{table*}

Table~\ref{UXI} reveals what has been already anticipated in subsect.~\ref{SSR}:
The original NLO interaction \cite{Haidenbauer:2015} with LECs fixed solely  
with the aim to meet the available experimental constraints on $\Xi^- p$ scattering 
leads to unrealistic strongly repulsive predictions for the $\Xi$ s.p. potential. 
However, the updated NLO interaction presented here makes clear that it is
possible to maintain agreement with those constraints and, at the
same time, achieve a moderately attractive $\Xi$-nuclear interaction. Evidently,
the values we obtain for $U_\Xi(0)$ are smaller than the commonly cited
benchmark based on the original fit to the BNL data \cite{Khaustov}. Yet, 
they are comparable to the results for the $\Xi N$ potentials ESC08c and fss2, 
for which likewise consistency with those BNL data is claimed. 

\begin{figure}[b!] 
\center{\includegraphics*[width=6.2cm]{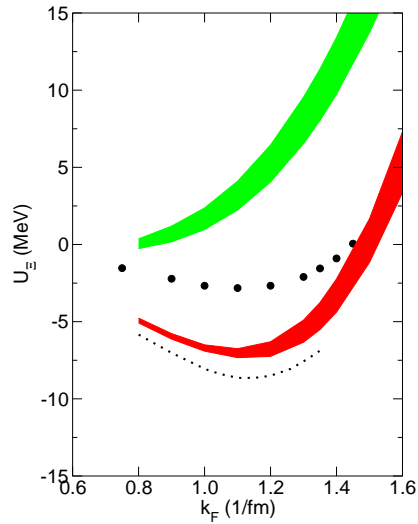}}
\caption{The $\Xi$ s.p. potential $U_\Xi(p_\Xi =0)$ as a function of the
Fermi momentum $k_F$. The black/red band shows the chiral EFT results to NLO for
variations of the cutoff in the range $\Lambda=500, ..., 650$~MeV. 
The outcome at LO is indicated by the grey/green band. 
The dotted line is the result for the Nijmegen potential ESC08c \cite{Rijken:2015}, 
the circles those for the quark-model potential fss2 \cite{Fujiwara07}, 
taken from Ref.~\cite{Kohno}. 
}
\label{kdep}
\end{figure}

As mentioned, for simplicity reasons we have performed the $G$-matrix 
calculation with the gap choice. However, we expect additional attraction in 
the order of $3$~MeV or more for the continuous choice, cf. the comments 
in subsect.~\ref{SSA}, which means that our results for $U_\Xi(0)$ are indeed 
very similar to those of the ESC08c potential. 
Nonetheless it is important to note that, contrary to that model, our EFT
interaction meets all the available empirical constraints on the 
$\Lambda\Lambda$ and $\Xi N$ interactions, see subsect.~\ref{SSR} and 
Ref.~\cite{Haidenbauer:2015}. Specifically, it does not lead to a 
near-threshold bound state in the $^3S_1$-$^3D_1$ partial wave
of the $\Xi N$ $I=1$ channel. The existence of such a state is practically
excluded by the mentioned experimental constraints \cite{Haidenbauer:2015}, 
and it is also not supported by the latest lattice QCD simulations close to 
the physical point \cite{Inoue:2018}. 

It should not be concealed that a considerable uncertainty in the predictions
for the $\Xi$ s.p.  potential comes from the contributions of the $P$-waves. 
Since there is very little
information on angular-dependent observables for $\La N$ and $\Si N$ 
and none for $\Xi N$, the pertinent LECs cannot be directly adjusted to data. 
Their values have been fixed in our works \cite{Haidenbauer:2013,Haidenbauer:2015} 
essentially by requiring that the contributions of the $P$-waves to the $\La p$ 
and $\Xi N$ cross sections remain small with increasing energy, to be in line with
the experiment. Of course, this prescription does not fix the sign, i.e. does not 
allow to determine whether the interactions in question are repulsive or attractive. 
We see from the results for $U_\Xi$ in Table~\ref{UXI} that in case of the original
NLO interaction about $25$~\% of the repulsion is generated by the $P$-waves. 
On the other hand, in case of the LO(600) interaction and the Nijmegen ESC08c
potential the $P$-wave contributions cancel more or less and the total result
is practically given by that of the $S$-waves alone. In other versions of the 
Nijmegen potential \cite{Yamamoto:2010} and also in other $\Xi N$ 
potentials \cite{Yamaguchi:2001} the $P$-waves provide even additional attraction 
in the order of $5$ MeV. Because of that in the updated NLO interaction
we have tried to achieve a balance between the $P$-wave contributions by 
readjusting some of the pertinent LECs, see Table~\ref{LEC} for details. In addition 
the sum of the $S$-wave contributions is listed separately in the table. Those are 
the ones which are primarily constrained by the available empirical information on the 
$\Xi N$ interaction. 

Predictions for the density dependence of the $\Xi$ s.p. potential are presented
in Fig.~\ref{kdep} where $U_\Xi (0)$ is shown as a function of $k_F$
for the updated NLO potential. Results for the LO interaction are shown too
and, for the ease of comparison, those for the ESC08c and fss2 potentials. 
Interestingly, the updated NLO interaction and the ESC08c potential exhibit a very similar 
density dependence. In particular, there is a trend towards repulsion with increasing 
density -- and in case of the NLO interaction the change of sign of $U_\Xi$
occurs already at rather moderate nuclear matter density. 
Accordingly, one expects that the onset density for the appearance of the $\Xi$ 
hyperons in neutron stars should be fairly high for such interactions \cite{Yamamoto:2016}, 
and, in turn, there should be a rather small effect on the equation-of-state 
\cite{Weissenborn:2012,Weissenborn:2012a}

\section{Summary}

We have investigated the in-medium properties of baryon-baryon potentials 
for the strangeness $S=-2$ sector, derived within chiral effective field theory 
up to next-to-leading order and constrained by available experimental 
information on the $\La\La$ and $\Xi N$ systems. 
In particular, we have evaluated the single-particle potential for the $\Xi$ 
hyperon in nuclear matter in a conventional $G$-matrix calculation. 

We could show that a $\Xi N$ interaction can be constructed which is in line
wth empirical constraints on the $\La\La$ $S$-wave scattering length and 
with published values and upper bounds for $\Xi^- p$ elastic and 
inelastic cross sections and which, at the same time, yields 
a moderately attractive $\Xi$-nuclear interaction as suggested by 
recent experimental evidence for the existence of $\Xi$-hypernuclei  
\cite{Nakazawa:2015,Nagae:2018}. 
The values obtained for the $\Xi$ single-particle potential $U_\Xi(0)$ in 
nuclear matter are with $-3$ to $-5$~MeV somewhat smaller than the commonly 
cited potential depth of $-14$~MeV. However, they are in line with results 
from comparable $G$-matrix calculations based on phenomenological $\Xi N$ 
potentials, where in applications to finite $\Xi$ systems 
consistency with data such as the spectrum in the $^{12}$C$(K^-,K^+)\,X$
reaction \cite{Khaustov} has been claimed \cite{Kohno:2010,Rijken:2015}.
It is also found that the original $\Xi N$ interaction published in 
Ref.~\cite{Haidenbauer:2015}, which was primarily meant to demonstrate that 
a baryon-baryon interaction fulfilling all experimental 
constraints on $\La\La$ and $\Xi N$ scattering can be established, 
is too repulsive in the medium. 

Of course, one has to acknowledge that the nuclear matter calculations we 
performed can provide only a first insight into the properties of the $\Xi N$ 
interaction in the medium. A more stringent test of those properties could 
be achieved by deducing effective interactions (from the $G$-matrix in nuclear 
matter) and applying them in calculations of finite hypernuclei.
Such a much more challenging investigation is postponed to the future. 

Finally, we want to emphasize that also for the updated NLO 
$\Xi N$ interaction introduced in the present paper no elaborate fine-tuning 
of the low-energy constants associated with the contact terms has been attempted.
For that one has to wait for further and more quantitative constraints from 
experiments involving the $\Xi$ hyperon and/or possibly employ \cite{Geng:2018} 
predictions from lattice QCD simulations \cite{Inoue:2018} once final and
confirmed results become available.
Further experimental information could come from future experiments involving 
the $\Xi$ hyperon at J-PARC \cite{JPARC} or FAIR \cite{Sanchez:2014}, 
but also from the study of pertinent correlations in heavy ion 
collisions or in high-energetic $pp$ scattering \cite{Fabbietti:2018}. 

\vskip 0.2cm \noindent
{\bf Acknowledgments} \\
This work is supported in part by the DFG and the NSFC through
funds provided to the Sino-German CRC 110 ``Symmetries and
the Emergence of Structure in QCD'' (DFG grant. no. TRR~110)
and the VolkswagenStiftung (grant no. 93562).
The work of UGM was supported in part by The Chinese Academy
of Sciences (CAS) President's International Fellowship Initiative (PIFI) (grant no.~2018DM0034).

\end{document}